\documentclass[fleqn,10pt]{wlscirep}
\usepackage[utf8]{inputenc}
\usepackage[T1]{fontenc}
\usepackage[final]{pdfpages}

\title{Learning physical properties of liquid crystals with deep convolutional neural networks}

\author[1]{Higor\ Y.\ D.\ Sigaki}
\author[2]{Ervin K. Lenzi}
\author[1,3]{Rafael S. Zola}
\author[4,5,6]{Matja{\v z} Perc}
\author[1,*]{Haroldo\ V.\ Ribeiro}

\affil[1]{Departamento de F\'isica, Universidade Estadual de Maring\'a, Maring\'a, PR 87020-900, Brazil}
\affil[2]{Departamento de F\'{\i}sica, Universidade Estadual de Ponta Grossa, Ponta Grossa, PR 84030-900, Brazil}
\affil[3]{Departamento de F\'{\i}sica, Universidade Tecnol\'ogica Federal do Paran\'a, Apucarana, PR 86812-460, Brazil}
\affil[4]{Faculty of Natural Sciences and Mathematics, University of Maribor, Koro{\v s}ka cesta 160, 2000 Maribor, Slovenia}
\affil[5]{Department of Medical Research, China Medical University Hospital, China Medical University, Taichung, Taiwan}
\affil[6]{Complexity Science Hub Vienna, Josefst{\"a}dterstra{\ss}e 39, 1080 Vienna, Austria}
\affil[*]{hvr@dfi.uem.br}

\begin{abstract}
Machine learning algorithms have been available since the 1990s, but it is much more recently that they have come into use also in the physical sciences. While these algorithms have already proven to be useful in uncovering new properties of materials and in simplifying experimental protocols, their usage in liquid crystals research is still limited. This is surprising because optical imaging techniques are often applied in this line of research, and it is precisely with images that machine learning algorithms have achieved major breakthroughs in recent years. Here we use convolutional neural networks to probe several properties of liquid crystals directly from their optical images and without using manual feature engineering. By optimizing simple architectures, we find that convolutional neural networks can predict physical properties of liquid crystals with exceptional accuracy. We show that these deep neural networks identify liquid crystal phases and predict the order parameter of simulated nematic liquid crystals almost perfectly. We also show that convolutional neural networks identify the pitch length of simulated samples of cholesteric liquid crystals and the sample temperature of an experimental liquid crystal with very high precision.
\end{abstract}
\begin{document}

\flushbottom
\maketitle

\thispagestyle{empty}

\section*{Introduction}

The idea of having a machine capable of imitating intelligent human behavior broadly defines the field of artificial intelligence. By quoting McCarthy, who first coined this term in 1956, we may define artificial intelligence as ``the science and engineering of making intelligent machines, especially intelligent computer programs''~\cite{mccarthy2007what}. In this context, machine learning can be understood as a subfield of artificial intelligence and represents a technique for realizing artificial intelligence. The first use of the term machine learning is usually attributed to Samuel~\cite{Samuel1959some} in a 1959 article, where he verified the possibility of programming a computer to learn how to play the game of checkers. It was also around the same time that Rosenblatt proposed the perceptron algorithm~\cite{rosenblatt1958perceptron}, often considered to be the first artificial neural network, for pattern and shape recognition. In spite of important developments such as the Vapnik-Chervonenkis theory~\cite{vapnik1998statistical}, it was only after the beginning of the 21st century that machine learning, and particularly the class of deep learning~\cite{lecun2015deep,schmidhuber2015deep,goodfellow2016deep} methods started to be used more widely in areas such as game playing~\cite{silver2016mastering,silver2017mastering}, natural language processing~\cite{cambria2014jumping}, history of art~\cite{sigaki2018history}, speech recognition~\cite{geoffrey2012deep}, medical diagnosis~\cite{kononenko2001machine,esteva2017dermatologist,kim2018performance,lindsey2018deep,diamant2019deep,lee2019generating,abdeltawab2019novel}, and computer vision~\cite{hartley2000multiple,szeliski2010computer,goodfellow2016deep}.

Despite the great improvement in several applications of machine learning algorithms, the process of extracting meaningful information from images, that is, to replicate what the human visual system can do, proved to be a more challenging task~\cite{szeliski2010computer, prince2012computer}. Convolutional neural networks are considered to be the state-of-the-art tool for analyzing image data, and they also have the great advantage of not requiring manual feature extraction from images. In particular, these deep neural networks use a hierarchical cascade of convolutions and non-linear functions that automatically learn representations and low-level features directly from input images~\cite{Chollet}. This is one of the reasons these deep convolutional neural networks are very good at identifying objects in images.

Indeed, an acclaimed example of success of deep learning algorithms is documented in the ImageNet Large Scale Visual Recognition Challenge~\cite{russakovsky2015ImageNet}, an annual competition among computer algorithms for large-scale image classification and object detection. The introduction of a deep neural network model (AlexNet) by Krizhevsky et al.~\cite{krizhevsky2012imagenet} in 2012 is considered the major breakthrough in the competition not only because the top-5 error rate was reduced from 26\% to 16.4\%, but mainly due to the fact that deep learning algorithms became the top contestants ever since~\cite{russakovsky2015ImageNet}. It was also a deep learning approach (ResNet) that surpassed human-level performance in the ImageNet data set for the first time in 2015~\cite{he2015delving,he2016deep}.

More recently, deep learning methods have also entered the toolbox of researchers working in physical sciences~\cite{carleo2019machine}, such as physics~\cite{baldi2014searching,mukund2017transient,dreissigacker2019deep}, chemistry~\cite{goh2017chemception,ma2018deep,ZhangPrediction2019}, and material sciences~\cite{butler2018machine,jha2018elemnet,ziletti2018insightful,wei2019machine,tshitoyan2019unsupervised}. The development and the spread of these machine learning methods in combination with the increasing availability of large data sets has been recognized as the ``fourth paradigm of science''~\cite{agrawal2016perspective} and also the ``fourth industrial revolution''~\cite{schwab2015fourth}, having great potential to significantly enhance the role of computational methods in applied and fundamental research. Despite this unparalleled surge of applications of machine learning methods in the physical sciences, there are still several areas that have taken little to no advantage of these approaches. That is, for example, also the case for liquid crystal~\cite{degennes}. This is somewhat surprising because, like many biological and other complex materials, the physical properties of liquid crystals are often probed by means of imaging techniques such as polarized optical microscope imaging~\cite{zola2013surface}.

Here we aim to fill this gap by demonstrating that deep convolutional neural networks are very efficient in the task of predicting properties of liquid crystal samples directly from their optical textures. By optimizing the architecture of simple convolutional networks, we find that our deep learning approach identifies almost perfectly the phase (nematic or isotropic) of simulated samples of nematic liquid crystals and predicts their order parameter with very high accuracy. We also show that these convolutional networks classify the pitch length of simulated cholesteric liquid crystals with virtually 100\% accuracy, and we identify the sample temperature of an experimental liquid crystal (E7 mixture) with impressive precision.

The rest of this work is organized as follows. In the next section, we present a brief description of convolutional neural networks and all the results concerning the classification and regression tasks. We then conclude with a summary of the presented results and a discussion of their potential implications. In the Methods section, we describe the process of generating the simulated textures, the procedures for obtaining the experimental textures, and we also provide details about the implementation of our deep learning framework.

\section*{Results}\label{sec:results}

{\it Convolutional neural networks.}
Before we start describing our results, we briefly introduce the conceptual framework underlying convolutional neural networks~\cite{lecun2015deep,schmidhuber2015deep,goodfellow2016deep}. These networks are a particular type of artificial neural network where the basic unit is a neuron or a node. The design of artificial neural networks is inspired by  biological neural networks concepts and consists of layers of neurons that are fully connected to each other via weight values. Each neuron receives input values from the previous layer, calculates the weighted sum of these values, adds a bias term, evaluates a non-linear function (activation function), and outputs the function value to the next layer. This process loosely mimics the behavior of biological neurons that fire under enough stimuli. The process of training an artificial neural network consists of adjusting weights and bias terms so that input signals yield the required output values provided by the training set. The updates of weights and bias are based on a loss function that quantifies the output error and on a process known as backpropagation. During this process, a gradient descent algorithm is used to iteratively update weights and bias to minimize the loss function, and each complete pass over the training data is called an epoch.

The main difference between standard neural networks and convolutional neural networks is the existence of convolutional layers. Differently from fully connected layers, neurons in convolutional layers receive inputs from small and spatially continuous regions of the previous layer. These windowed inputs are further multiplied by filters that share weight across the entire input data. Thus, convolutional networks preserve the spatial structure and optimize filter weights that are responsible for extracting and detecting low-level features in different locations of the input data (usually images). To formally define a convolutional layer, we need to specify the size of the spatial windows (filter size) and the overlap between adjacent windows (stride). For instance, a filter size of $2\times2$ and a stride of $1$ means that the filter operates over windows with dimensions of $2\times2$ pixels (if the input is an image) that move in unitary steps over the input data. In addition to convolutional layers, convolutional networks usually have pooling or downsampling layers. A pooling layer operates similarly to convolutional layers, but instead of calculating weighted sums, it outputs simple calculations for each region such as maximum (max pooling) or average values (average pooling). These pooling layers summarize the presence of features, help in making feature representations invariant to small translations in the input data and reduce data dimensions (and amount of parameters), which in turn improves the computational efficiency of the network.

In our applications, the input data of the convolutional neural networks are texture images of liquid crystals, and the output are particular properties of these materials (phase state, average order parameter, pitch length, and sample temperature). The data set used here is basically the same as in Ref.~\cite{sigaki2019estimating} and comprises texture images obtained from simulated nematic and cholesteric liquid crystal samples as well as experimental textures obtained from E7 liquid crystal samples (a material commonly used in liquid crystal research). In Methods, we provide further details about the procedures involved in building this data set.

There is a myriad of possibilities for design choices of convolutional neural network architectures (number of layers, number and size of filters, strides, and so on). These choices are mostly empirical, dependent on the type of input data, and often inspired by other architectures that proved to be successful at particular tasks. However, some design patterns are common to several network architectures~\cite{smith2016deep}. These include parsimony, symmetry, incremental feature construction, and downsampling strategy as we go deeper into the network~\cite{smith2016deep}. Our particular design choices have been guided by these principles but are also based on trial-and-error procedures as well as cross-validation over a few network parameters.

\begin{figure*}[!ht]
\centering
\includegraphics[width=0.98\textwidth, keepaspectratio]{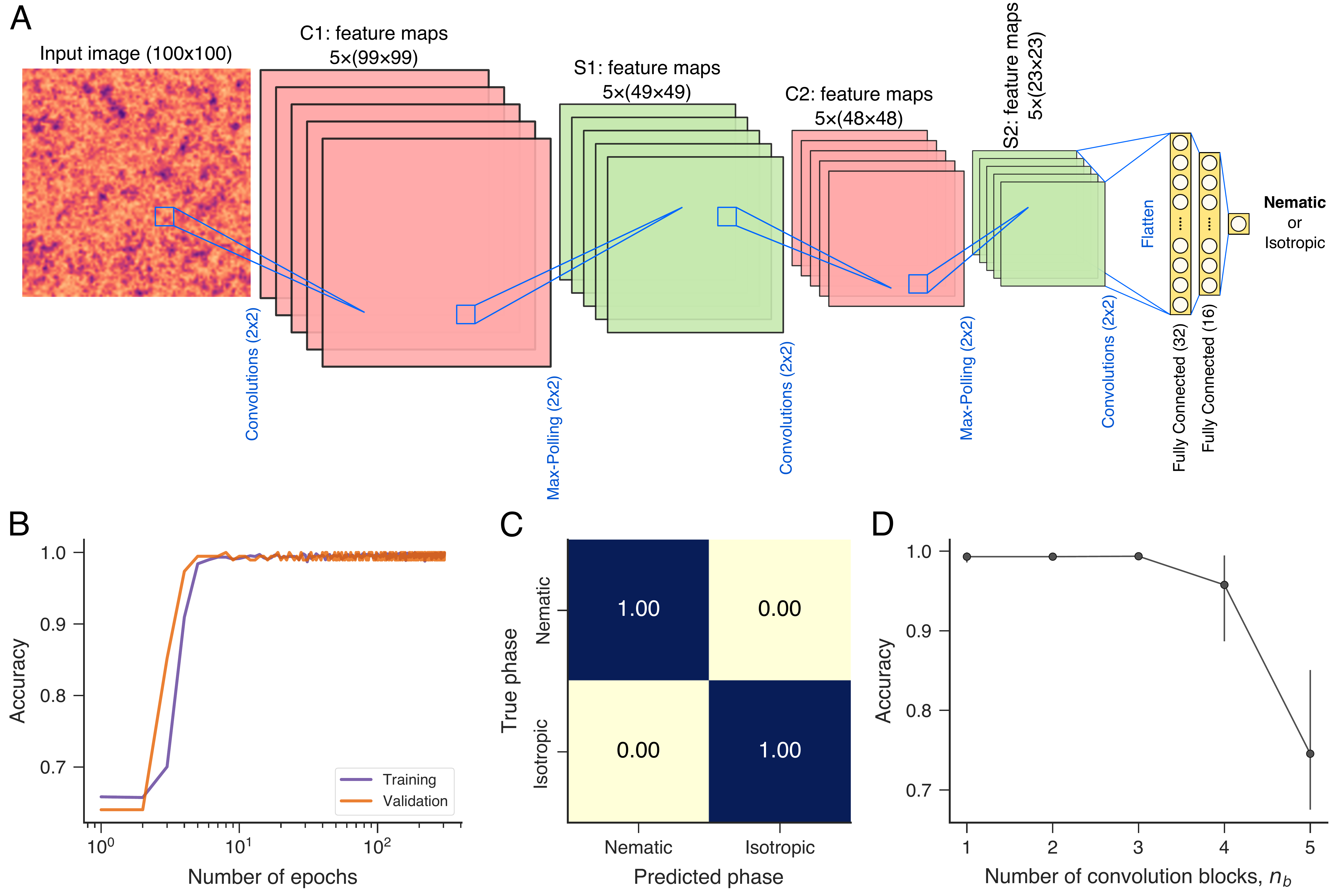}
\caption{{\bf Predicting the phase of liquid crystals with convolutional neural networks.} (A) Schematic representation of the network architecture used for predicting the phase of liquid crystals from their textures. This network comprises two blocks of convolutional (red) and max-pooling (green) layers followed by two fully connected layers (yellow) and an output layer. An input image of size $100\times100$ pixels is convolved with five $2\times2$ filters (with unitary strides), yielding five $99\times99$ feature maps (C1 in red) that are passed through rectified linear unit (ReLU) activation functions. These feature maps are then passed to $2\times2$ max-pooling operations that coarse grain the representation to five $49\times49$ feature maps (S1 in green). Next, these feature maps pass to the same configuration of convolution and max-pooling blocks, yielding five $23\times23$ feature maps (S2 in red) that are flattened and passed through two fully connected layers with 32 and 16 nodes. Finally, the phase classification (nematic or isotropic) takes place in the output layer via a sigmoid activation function (corresponding to logistic regression). (B) Training and validation scores (fraction of correct classifications) as a function of the number of epochs used during the training stage. We separate 15\% of data as test set, and the remaining is divided into training (80\%) and validation (20\%) sets (all obtained in a stratified manner). (C) Confusion matrix obtained when applying the trained network to the test set (never exposed to the trained network). (D) Accuracy of the network over the test set as a function of the number of convolution (and max-pooling) blocks $n_b$ in the architecture (panel A corresponds to $n_b=2$). The circles are average values over ten realizations of the training procedures, and the error bars are 95\% confidence intervals.}
\label{fig:1}
\end{figure*}

{\it Predicting the phase of liquid crystals.}
In a first application, we use a convolutional neural network to detect whether a nematic liquid crystal is in the nematic or isotropic phase. To do so, we numerically generate textures from a model (see Methods) presenting nematic to isotropic transition at a critical temperature $T_c$. Textures with temperature below $T_c$ are labeled as ``nematic'' and those with temperature above $T_c$ are considered as ``isotropic''. This classification task is visually straightforward when textures are obtained from temperatures far from the critical temperature but becomes challenging with textures around the critical temperature~\cite{sigaki2019estimating}. Figure~\ref{fig:1}A illustrates the network architecture initially used for this task. In this network, input images (100$\times$100 pixels) pass through two blocks of 2$\times$2 convolutions and 2$\times$2 max-pooling layers, followed by two fully connected layers (with 32 and 16 nodes, respectively) and an output layer. We use rectified linear unit (ReLU) activation functions in all convolutional and fully connected layers, while the output layer uses a sigmoid activation function (corresponding to logistic regression).

We separate 15\% of data for final evaluation (test set) of the model and use the remaining as validation (20\%) and training (80\%) sets. The network parameters are optimized using the Adam algorithm~\cite{kingma2014adam} (learning rate of 0.001), and the loss function is the binary cross-entropy (commonly used in binary classification). To avoid overfitting, we apply an early stopping regularization procedure (with patience set to 10 epochs) and an L2 weight regularization (hyperparameter $\lambda=0.001$) over all convolutional and fully connected layers. Figure~\ref{fig:1}B depicts the training and validation scores (fraction of correct classifications) as a function of the number of epochs, where we note that this network achieves ideal accuracy with just a few training epochs. Figure~\ref{fig:1}C shows the confusion matrix obtained by applying the trained network to the 15\% of data never exposed to the algorithm. These results indicate that our network also achieves perfect accuracy in identifying the liquid crystal phase (nematic or isotropic) in the test set.

We also test if variations of the network architecture shown in Figure~\ref{fig:1}A are capable of classifying phases with similar performance. To do so, we consider network variations where only the number of convolution blocks $n_b$ (followed by max-polling layers) changes from 1 to 5 (the network of Figure~\ref{fig:1}A corresponds to $n_b=2$). We thus train ten realizations of each of these networks by using the same procedure described for the architecture of Figure~\ref{fig:1}A. After training, we estimate the average accuracy of the classification task in the test set as a function of $n_b$. The results of Figure~\ref{fig:1}D indicate that networks with $n_b=1,2$, or 3 convolution blocks are equally good at classifying liquid crystal phases with accuracy very close to the ideal value. Thus, it would be preferable to use $n_b=3$ when deploying this model in a more practical application, since the number of fitting parameters diminishes when the number of convolution blocks increases, which in turn facilitates the training procedures. We further observe in Figure~\ref{fig:1}D that the classification performance decreases substantially when increasing the number of convolution blocks beyond $n_b=3$, reaching an accuracy of $\sim0.7$ for $n_b=5$. 

\begin{figure*}[!ht]
\centering
\includegraphics[width=0.98\textwidth, keepaspectratio]{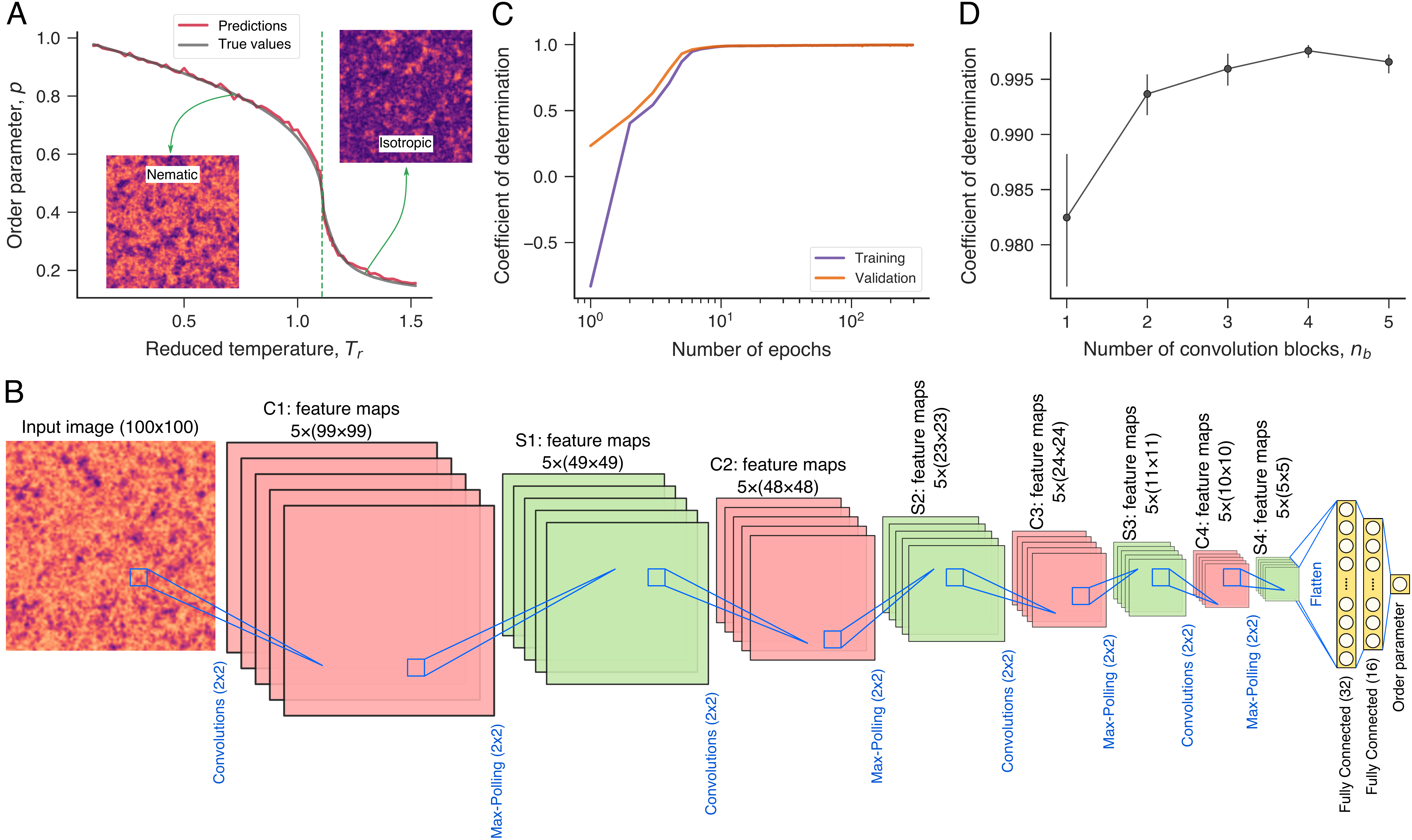}
\caption{{\bf Predicting the order parameter of liquid crystals with convolutional neural networks.} (A) Dependence of the order parameter $p$ on the reduced temperature $T_r$ for a simulated nematic liquid crystal (gray line). The vertical dashed line indicates the critical temperature $T_c=1.1075$ separating the nematic ($T_r<T_c$) and isotropic ($T_r>T_c$) phases. The insets show typical textures of each phase. (B) Schematic representation of the network architecture used for the regression task of predicting the order parameter $p$ from the textures. This network has the same general structure used for phase classification and comprises four blocks of convolutional (red) and max-pooling (green) layers followed by two fully connected layers (yellow) and an output layer. The only difference is in the last layer, where we use a linear activation function for estimating the order parameter $p$. (C) Coefficient of determination (between actual and predicted values) estimated from the training and validation sets as a function of the number of epochs used during the training stage. We separate 15\% of data as test set, and the remaining is divided into training (80\%) and validation (20\%) sets (all obtained in a stratified manner). The trained network yields a coefficient of determination of $\approx0.997$ when applied to the test set, and the red line in panel A illustrates the accuracy of the network predictions. (D) Coefficient of determination obtained from the test set as a function of the number of convolution (and max-pooling) blocks $n_b$ in the architecture (panel B corresponds to $n_b=4$). The circles are average values over ten realizations of the training procedures, and the error bars are 95\% confidence intervals.}
\label{fig:2}
\end{figure*}

{\it Predicting the order parameter of liquid crystals.}
In another application, we propose to predict the order parameter $p$ of simulated liquid crystals directly from their textures. The order parameter describes the orientational order of a sample; it is considered one of the most important physical properties of the nematic phase since other anisotropic properties are determined from $p$. Figure~\ref{fig:2}A shows the dependence of $p$ on the temperature $T_r$, where we observe that $p$ decreases with $T_r$. This liquid crystal undergoes a transition from nematic to isotropic phase when the temperature exceeds the critical value $T_c= 1.1075$~\cite{sigaki2019estimating}. Differently from the phase classification, we now have a regression problem where the network output is a continuous number representing the order parameter $p$. For this regression task, we consider essentially the same network architectures used for classifying liquid crystal phases. We only replace the sigmoid activation function of the output layer by a linear activation function. Figure~\ref{fig:2}B depicts the architecture with four convolutional (and max-pooling) layers ($n_b=4$).

We train this network by optimizing the mean square error (loss function) and following the same procedures used for the phase classification. Figure~\ref{fig:2}C shows that the coefficients of determination (between actual and predicted values) for training and validation sets approach $1$ after only a few training epochs. We also find a coefficient of determination of $\approx0.997$ when applying the trained network to the test set. This result demonstrates that our convolutional neural network is remarkably efficient in predicting the order parameter $p$, outperforming a shallow learning approach based on two image features (permutation entropy and statistical complexity) and $k$-nearest neighbors algorithm~\cite{sigaki2019estimating}. Figure~\ref{fig:2}A shows a comparison between actual and predicted values for the order parameter $p$, where we visually observe the high degree of accuracy achieved by the network.

We further investigate how the number of convolution (and max-pooling) blocks $n_b$ affects network accuracy. To do so, we train ten realizations of the network for a given value of $n_b\in\{1,2,3,4,5\}$ and estimate the average value of the coefficient of determination for the test set. Figure~\ref{fig:2}D shows that these networks display excellent precision for different number of convolution blocks, but an optimal performance occurs for $n_b=4$ (the architecture of Figure~\ref{fig:2}A).

\begin{figure*}[!ht]
\centering
\includegraphics[width=0.98\textwidth, keepaspectratio]{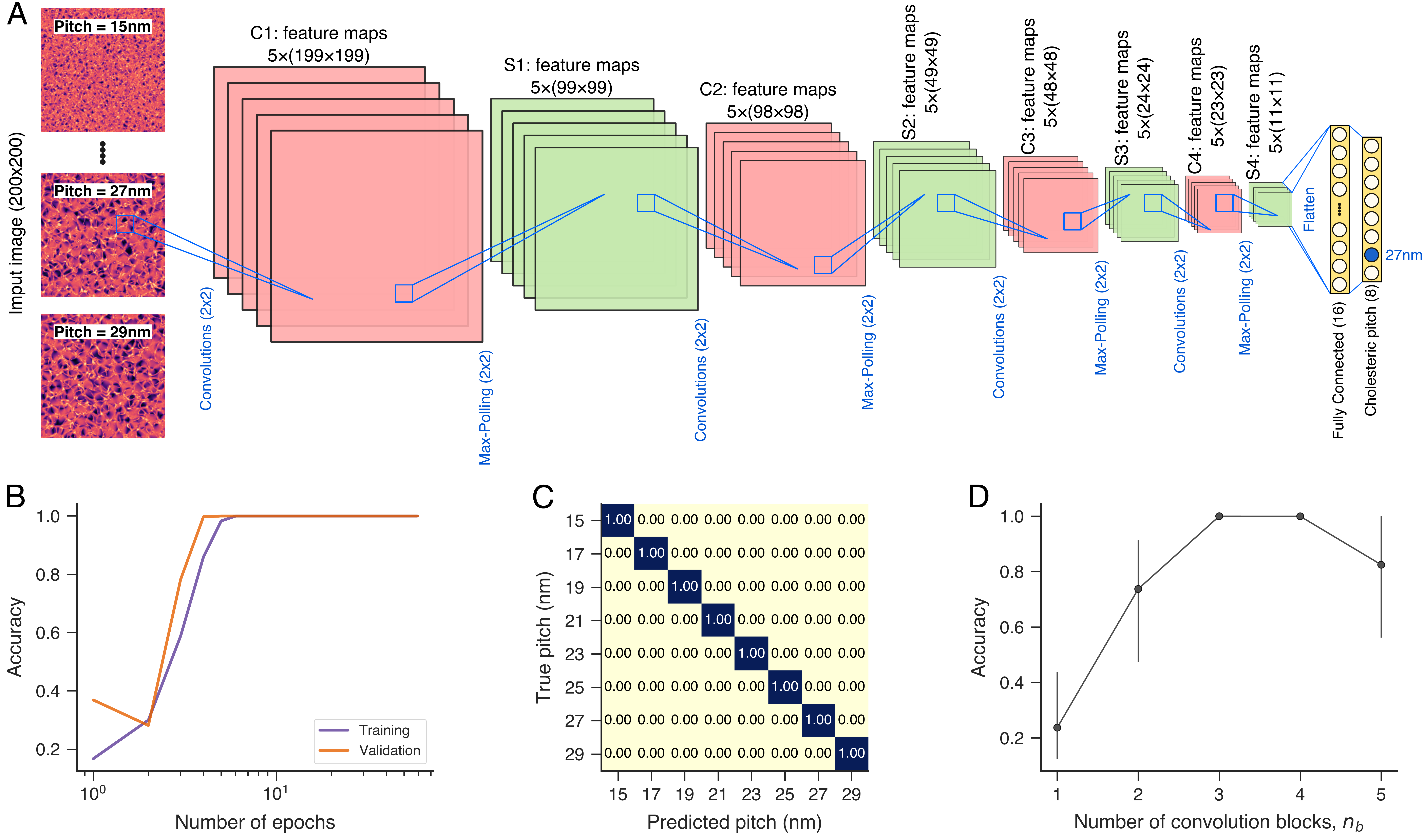}
\caption{{\bf Predicting the pitch length of cholesteric liquid crystals with convolutional neural networks.} (A) Illustration of the network architecture used for classifying the pitch values. This network has the same general structure used for classifying phases and predicting the order parameter. The difference is in the last layer that is now composed of 8 nodes with softmax activation functions. (B) Training and validation scores (accuracy, the fraction of correct classifications) as a function of the number of training epochs. We use 15\% of data as test set and the remaining is divided into training (80\%) and validation (20\%) sets (all obtained in a stratified manner). (C) Confusion matrix obtained by applying the trained network to the test set. The diagonal form shows that the trained network achieves a perfect classification of all pitch values in the test set. (D) Accuracy estimated from the test set as a function of the number of convolution (and max-pooling) blocks $n_b$ in the architecture (panel A corresponds to $n_b=4$). The markers represent the average values obtained from ten realizations of the training procedures, and the error bars are 95\% confidence intervals.}
\label{fig:3}
\end{figure*}

{\it Predicting the pitch length of cholesteric liquid crystals.}
Cholesteric liquid crystals are materials displaying a periodical helical (chiral) structure. This arrangement might be viewed as if formed by layers in between which the preferential director axis changes periodically with the period known as pitch length $\eta$. The value of $\eta$ is easily estimated when the helical axis is perpendicular to the viewing direction of an optical microscope but cannot be obtained from standard experimental arrangements (Grandjean textures) used in reflective displays~\cite{zheng2017controllable}. It can thus be of practical interest to find a simple way of estimating the pitch $\eta$ directly from textures of cholesteric liquid crystals. To test whether convolutional neural networks are useful for predicting the value of $\eta$, we have built a data set of textures associated with different values of $\eta\in\{15,17,19,\dots,29\}$~nm from numerical simulations (see Methods). We then apply the same general network architecture used with nematic textures for the task of classifying the values of $\eta$. Figure~\ref{fig:3}A shows a network with $n_b=4$ convolution blocks (and max-polling) used with the cholesteric textures. When compared with networks used with nematic textures, the only difference is in the last layer that now comprises 8 nodes (one for each pitch value) with softmax activation functions (commonly used in multiclass classification tasks). We train this network by following the same procedures used before and by considering the categorical cross-entropy as the loss function. Figure~\ref{fig:3}B shows that the training and validation scores approach the ideal accuracy after about 10 training epochs. Figure~\ref{fig:3}C further demonstrates the high accuracy of this network by depicting the confusion matrix estimated from the test set (15\% of data never presented to the algorithm). We note that this network perfectly classifies all pitch values. This performance is quite superior to one obtained by the shallow learning approach reported in Ref.~\cite{sigaki2019estimating}, where an accuracy of $\approx85\%$ is reported. We also investigate the accuracy of different network architectures by changing the number of convolution blocks $n_b$. Results of Figure~\ref{fig:3}D shows that the average accuracy is very low for $n_b<3$, reaches an optimum value for $n_b=3$ and $4$, and decreases when $n_b=5$.

\begin{figure*}[!ht]
\centering
\includegraphics[width=0.98\textwidth, keepaspectratio]{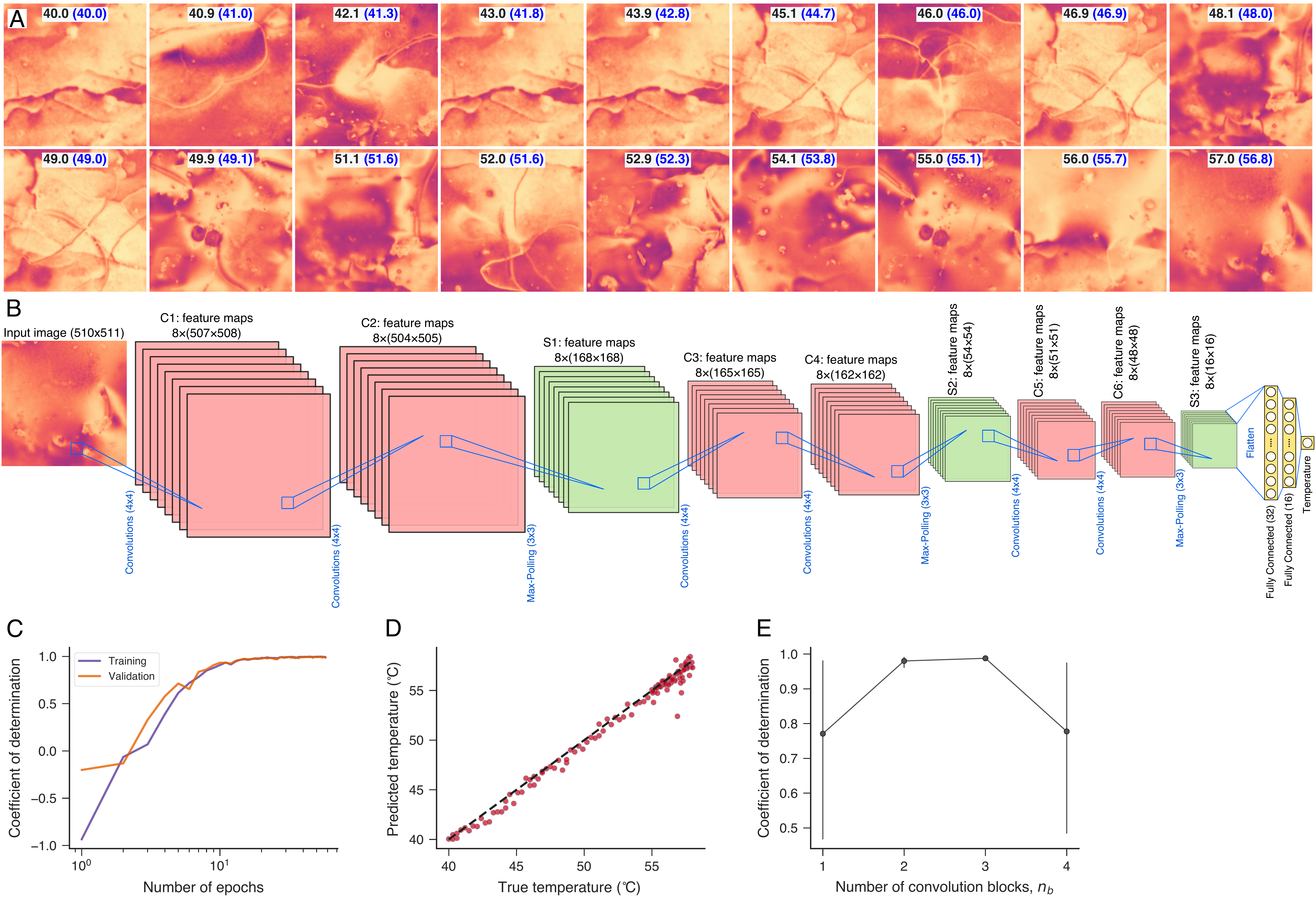}
\caption{{\bf Predicting the sample temperature of E7 liquid crystals with convolutional neural networks.} (A) Examples of experimental textures obtained from polarized optical microscope imaging for different samples at different temperatures (indicated within the images in degree Celsius). (B) Schematic representation of the network architecture used for the regression task of predicting the sample temperature. This network architecture is slightly different from all others we have used so far; it is composed of three blocks of eight 4$\times$4 convolutions followed by other eight 4$\times$4 convolutions and by eight 3$\times$3 max-poling layers. After the last max-poling operations (S3), we have two fully connected layers (with 32 and 16 nodes) and an output layer with a single node and a linear activation function. All convolutional layers use a ReLU activation function. (C) Coefficient of determination estimated from the training and validation sets as a function of the number of training epochs. We separate 15\% of data as test set and the remaining is divided into training (80\%) and validation (20\%) sets (all obtained in a stratified manner). The coefficient of determination calculated for the test set is $\approx0.982$. (D) Relationship between predicted and true temperature values obtained by applying the training network to the test set (the dashed line is the 1:1 relationship). The values between brackets in Panel A also indicate network predictions. (E) Coefficient of determination estimated from the test set as a function of the number of convolution (and max-pooling) blocks $n_b$ in the architecture (panel B corresponds to $n_b = 3$). The markers represent the average values obtained from ten realizations of the training procedures, and the error bars are 95\% confidence intervals.
}
\label{fig:4}
\end{figure*}

{\it Predicting the sample temperature of E7 liquid crystals.}
In a last application, we propose to predict the sample temperature from experimental textures of E7 liquid crystals. These materials are a multicomponent mixture (cyanobiphenyl and cyanoterphenol) frequently used for the production of commercial displays. E7 liquid crystals exhibit a nematic to isotropic transition when the temperature exceeds the critical value of $\approx58^\circ$~C\cite{sigaki2019estimating}. Figure~\ref{fig:4}A shows examples of nematic textures obtained at different temperatures. We have initially verified that the general architecture used in all previous applications does not yield good results when dealing with these experimental textures. Because of that, we propose to slightly modify the network architecture by including additional convolutional layers before each max-polling operation. We also increase the number and size of the convolution filters (there are now eight $4\times4$ filters per convolution block) as well as the size of the max-polling filters (that are now $3\times3$ pixels). The fully connected layers remain equal to the previous cases, that is, we have two fully connected layers with 32 and 16 nodes, followed by an output layer. ReLU activation functions are used after all convolution operations, and a linear activation function is used in the output layer. Figure~\ref{fig:4}B illustrates the modified network structure with $n_b=3$ convolution blocks.

In spite of the modification in the network architecture, the training and regularization procedures remain the same. We also use the mean square error as the loss function for this regression problem. Figure~\ref{fig:4}C shows the coefficient of determination for the training and validation sets as a function of the training epochs. We observe that both scores approach the ideal value after a few training epochs. The results of Figure~\ref{fig:4}D shows the relationship between predicted and true temperatures obtained by applying the trained network to the test set. This relationship closely follows the 1:1 line (dashed line) and has a coefficient of determination of $\approx0.982$, indicating the high precision achieved by our approach. Figure~\ref{fig:4}A also shows a comparison between the actual temperature values associated with each texture and the network predictions (values within brackets). The accuracy of our network outperforms the shallow learning approach reported in Ref.~\cite{sigaki2019estimating}, where a coefficient of determination of $\approx0.93$ is obtained with the $k$-nearest neighbors algorithm. We have also investigated the accuracy of our approach with different number of convolution blocks $n_b$. Figure~\ref{fig:4}E shows the average coefficient of determination as a function of $n_b$, where we notice that the optimal accuracy occurs for $n_b=2$ or $n_b=3$.

\section*{Discussion}

While neural networks and other machine learning methods will not entirely replace experimental procedures, these methods are already improving experimental results and overcoming several difficulties in experimental analysis. From the results shown here, we expect that the use of neural networks in experimental situations may experience a large growth in the near future. In fact, there are several basic and applied science scenarios that can benefit from these techniques, which goes far beyond the exclusive use in liquid crystals research. 

Upon correct training and design, neural networks and other machine learning methods have proved useful for identifying phase transitions~\cite{cao2018machine,sigaki2019estimating} (thus coming in aid to regular thermal characterization such as differential scanning calorimetry measurements), pattern formation and pattern identification~\cite{minor2020end} (thus replacing the need for ever more specialized imaging tools and helping researchers to pick up very small details). Other examples of success of machine learning methods include investigation of biological materials~\cite{diaz2020vitro}, deep space investigation~\cite{shallue2018identifying}, and the search for early stage breast cancer by looking for calcification cluster on mammograms~\cite{cai2019breast}. It is likely that, in the foreseeable future, machine learning (particularly convolutional neural networks) will find its use in basically all imaging tools: from optical to electronic microscopy. 

Our work have demonstrated the usefulness of deep convolutional neural networks for predicting physical properties of liquid crystals directly from their optical textures. We worked out a series of applications with simulated and experimental textures, in which these networks showed to be quite efficient for predicting phases (nematic or isotropic), order parameters, cholesteric pitches, and sample temperatures of different liquid crystals. Our results thus help reducing the shortage of machine learning research, and in particular the application of deep learning algorithms, on liquid crystals.



\section*{Methods}

\subsection*{Implementing neural networks}

All convolutional neural networks used here are implemented in Python language via TensorFlow~\cite{chollet2015keras} with the Keras~\cite{tensorflow2015-whitepaper} high-level API. We have particularly used the Keras sequential model, where deep neural networks are created by sequentially assembling layers. The general network architectures used in all applications are detailed in Figures~\ref{fig:1}A, \ref{fig:2}B, \ref{fig:3}A, and ~\ref{fig:4}B. Except for the study with E7 textures, all networks are built by stacking $n_b$ blocks of convolutional layers followed by max-polling layers. Next, the resulting feature maps are flattened and passed to two fully connected layers with respectively 32 and 16 nodes. A ReLU activation function is used in all convolutional and fully connected layers. The output layer is composed of a single node with a linear activation function in the regression tasks (when predicting the order parameter and sample temperature). The output has a single node with a sigmoid activation function in the binary (nematic or isotropic) classification of phases, and eight nodes with softmax activation functions when classifying the cholesteric pitches ($\eta\in\{15,17,19,\dots,29\}$~nm). For the E7 textures, the network architecture is modified and comprises $n_b$ blocks of convolutional layers followed by other convolutional layers followed by max-polling layers. The resulting feature maps are then passed through the same structure of fully connected layers.

We train these networks by optimizing a loss function with the Adam stochastic optimization algorithm~\cite{kingma2014adam}. In all applications, we have fixed the learning rate in 0.001 and the exponential decay rates in $\beta_1=0.9$ and $\beta_2 = 0.999$ (commonly used settings~\cite{kingma2014adam,chollet2015keras}). We use the mean square error as the loss function for the regression tasks (predictions of order parameters and temperatures), and binary and categorical cross-entropy for the classification tasks (predictions of phases and pitches, respectively). In all applications, we separate 20\% of data for final validation of the model (test set) and divide the remaining of data into training (80\%) and validation (20\%) sets. To avoid overfitting, we consider an early stopping regularization procedure that ends the training process when the validation loss function stops improving within a ten epochs interval (the patience parameter). We have also included a penalty term in the loss function proportional to the sum of the squares of the layer parameters (an L2 norm for regularization, where $\lambda=0.005$ is the constant of proportionality). This regularization procedure also helps in avoiding overfitting and reduces fluctuations in the loss function. In the Supplementary Information we show excerpt of codes used for defining the convolutional neural networks in our work (Codes~1, 2, 3 and 4) and also a minimal example of code used for training these networks (Code 5).

\subsection*{Data set of liquid crystal textures}

The data set used in our applications comprises two different types of textures obtained from numerical simulations and experimental textures obtained from polarized optical microscope imaging. These experimental and simulated textures are basically the same reported in Ref.~\cite{sigaki2019estimating}, where further details can be found. Here we shall briefly describe the approach.

{\it Nematic and isotropic textures from Monte Carlo simulations.}
The textures used for predicting the liquid crystal phase and order parameters are obtained by Monte Carlo simulations of the so-called Lebwohl-Lasher model~\cite{lebwohl1972nematic}. This model describes headless spins located over the sites of a $n_x\times n_y\times n_z$ lattice (with $n_x=n_y=100$ and $n_z=20$). Unit vectors representing their directions characterize each of these spins. These spins interact with each other via the Lebwohl-Lasher potential $\Phi_{ij} \propto \cos \vec{u}_i\cdot\vec{u}_j$, where $\vec{u}_i$ and $\vec{u}_i$ refers to $i$-th and $j$-th spins. We use periodic boundary conditions along the $x$ and $y$ directions, while the first and last layer of spins along the $z$ direction have a fixed direction pointing along the $y$ and $x$ directions, respectively. We simulate this system with a given reduced temperature $T_r$, discarding the initial $10^4$ Monte Carlo steps to avoid transient behaviors. We use other $10^4$ steps for estimating the average order parameter $p$ and the textures are obtained with the Stokes-Muller methodology~\cite{berggren1994computer} averaged over the latest 50
Monte Carlo steps. We run 200 realizations for each value of $T_r\in\{0.10,0.12,0.14,\dots,1.52\}$, yielding 14,400 images of size $100\times100$ pixels. This model presents nematic to isotropic transition at the critical temperature $T_c = 1.1075$~\cite{sigaki2019estimating}, so that textures with $T_r<T_c$ are labeled as ``nematic'' and those with $T_r>T_c$ are considered ``isotropic''. An average order parameter $p$ is also associated with each texture.

{\it Cholesteric textures from simulations.}
The cholesteric textures used for classifying the pitches are obtained from simulations of the Landau-de Gennes modeling approach~\cite{Ravnik2009Landau} (continuum elastic theory). These simulations are carried out via finite differences method in a uniform grid of size $200\times200\time20$, and the liquid crystal parameters are chosen to mimic the 5CB liquid crystal~\cite{degennes}. This system is simulated with periodic boundary conditions and for different values of the pitch $\eta\in\{15,17,19,\dots,29\}$~nm. The optical textures are estimated via the Jones $2\times2$ method~\cite{Wu2006}, an approach that is well known to produce textures very similar to experimental results~\cite{Sec2012,sigaki2019estimating}. We generate 1,000 textures for each value of $\eta$, yielding 8,000 images of size $200\times200$ pixels.

{\it Experimental E7 textures.}
The experimental textures of E7 liquid crystals are obtained via polarized optical microscope imaging. Each sample consists of a rectangular capillary (300~$\mu$m $\times$ 4~mm) filled with the E7 mixture. We place these samples under a polarized optical microscope setup coupled with a temperature controller. We take pictures of the textures every 90~s, starting at 40~$^\circ$C and heating the sample at a constant rate of 0.2~$^\circ$C per minute up to 58~$^\circ$C (sample temperatures are thus $T\in\{40,40.3,40.6,\dots,58\}~^\circ$C). All image files have dimensions $2047\times1532$ pixels (with 24 bits per pixel) and are converted into grayscale via luminance transformation~\cite{scikit-image}. These images are sliced into 12 non-overlapping parts of size 510$\times$511 pixels. We further use an augmentation procedure that adds a copy of each image horizontally and vertically flipped to the data set. We consider 5 different samples, yielding 180 images per temperature value (total of 10980 images).

\bibliography{cnn_liquid_crystal}

\section*{Acknowledgements}

This research was supported by Coordena\c{c}\~ao de Aperfeicoamento de Pessoal de N\'ivel Superior (CAPES) and Conselho Nacional de Desenvolvimento Cient\'ifico e Tecnol\'ogico (CNPq). E.K.L. acknowledges the support of National Institutes of Science and Technology of Complex Systems (INCT-SC). R.S.Z. thanks the National Institute of Science and Technology Complex Fluids (INCT-FCx), and the S\~ao Paulo Research Foundation (FAPESP, Grant No. 2014/50983-3). M.P. acknowledges financial support of the Slovenian Research Agency (Grant Nos. J4-9302, J1-9112, and P1-0403). H.V.R. thanks for the financial support of CNPq (Grant Nos. 407690/2018-2 and 303121/2018-1).

\section*{Data availability}
All data supporting the findings of this study are available from the corresponding author on reasonable request.

\section*{Author contributions statement}

H.Y.D.S., E.K.L., R.S.Z., M.P., and H.V.R. designed research, performed research, analyzed data, and wrote the paper.

\section*{Additional information}

\textbf{Competing interests:} The authors declare that they have no conflict of interest. 



\section*{Supplementary material}

# Learning physical properties of liquid crystals with deep convolutional neural networks

Higor Y. D. Sigaki et al., Sci. Rep., 2020.

## DEFINING THE CONVOLUTIONAL NEURAL NETWORKS OF OUR WORK WITH HIGH-LEVEL API

```python
import tensorflow as tf
from tensorflow.keras import layers
from tensorflow import keras

def keras_model_for_predicting_phase(nblocks=2, nfilters=5,
                                     filter_size=2, polling_size=2,
                                     dense=[32, 16], reg_lamda=0.005):
    """
    Returns a CNN used for predicting the phase of nematic liquid crystals.

    Parameters:
    nblocks: number of convolution and max-pooling blocks.
    nfilters: number of convolution filters in each convolution block.
    filter_size: size of the convolution filters.
    polling_size: size of the pooling filters.
    dense: a list with the number of neurons in each dense layer.
    reg_lamda: L2 norm for regularization.

    Returns:
    A tf.keras model.
    """

    model = tf.keras.Sequential()
    model.add(
        layers.Conv2D(nfilters,
                      filter_size,
                      activation='relu',
                      input_shape=(100,100,1,),
                      kernel_regularizer=keras.regularizers.l2(reg_lamda)))
    model.add(layers.MaxPooling2D(pool_size=(polling_size, polling_size)))

    for i in range(nblocks - 1):
        model.add(
            layers.Conv2D(nfilters, filter_size, activation='relu',
                          kernel_regularizer=keras.regularizers.l2(reg_lamda)))
        model.add(layers.MaxPooling2D(pool_size=(polling_size, polling_size)))

    model.add(layers.Flatten())

    for ndense in dense:
        model.add(
            layers.Dense(ndense, activation='relu',
                         kernel_regularizer=keras.regularizers.l2(reg_lamda)))

    model.add(layers.Dense(1, activation='sigmoid'))

    return model
```

Code 1. Python code used for defining a function that returns a keras convolutional neural network model for predicting the phase of nematic liquid crystals.

```python
import tensorflow as tf
from tensorflow.keras import layers
from tensorflow import keras

def keras_model_for_predicting_order_parameter(nblocks=4,
                                               nfilters=5,
                                               filter_size=2,
                                               polling_size=2,
                                               dense=[32, 16],
                                               reg_lamda=0.005):
    """
    Returns a CNN used for predicting the order parameter of nematic liquid crystals.

    Parameters:
    nblocks: number of convolution and max-pooling blocks.
    nfilters: number of convolution filters in each convolution block.
    filter_size: size of the convolution filters.
    polling_size: size of the pooling filters.
    dense: a list with the number of neurons in each dense layer.
    reg_lamda: L2 norm for regularization.

    Returns:
    A tf.keras model.

    """
    model = tf.keras.Sequential()
    model.add(
        layers.Conv2D(nfilters,
                      filter_size,
                      activation='relu',
                      input_shape=(100,100,1,),
                      kernel_regularizer=keras.regularizers.l2(reg_lamda)))
    model.add(layers.MaxPooling2D(pool_size=(polling_size, polling_size)))

    for i in range(nblocks - 1):
        model.add(
            layers.Conv2D(nfilters,
                          filter_size,
                          activation='relu',
                          kernel_regularizer=keras.regularizers.l2(reg_lamda)))
        model.add(layers.MaxPooling2D(pool_size=(polling_size, polling_size)))

    model.add(layers.Flatten())

    for ndense in dense:
        model.add(
            layers.Dense(ndense,
                         activation='relu',
                         kernel_regularizer=keras.regularizers.l2(reg_lamda)))

    model.add(layers.Dense(1, activation='linear'))

    return model
```

Code 2. Python code used for defining a function that returns a keras convolutional neural network model for predicting the order parameter of nematic liquid crystals.

```python
import tensorflow as tf
from tensorflow.keras import layers
from tensorflow import keras

def keras_model_for_predicting_pitch(nblocks=4,
                                     nfilters=5,
                                     filter_size=2,
                                     polling_size=2,
                                     dense=[16],
                                     reg_lamda=0.005):
    """
    Returns a CNN used for predicting the pitch of cholesteric liquid crystals.

    Parameters:
    nblocks: number of convolution and max-pooling blocks.
    nfilters: number of convolution filters in each convolution block.
    filter_size: size of the convolution filters.
    polling_size: size of the pooling filters.
    dense: a list with the number of neurons in each dense layer.
    reg_lamda: L2 norm for regularization.

    Returns:
    A tf.keras model.

    """
    model = tf.keras.Sequential()
    model.add(
        layers.Conv2D(nfilters,
                      filter_size,
                      activation='relu',
                      input_shape=(200,200,1,),
                      kernel_regularizer=keras.regularizers.l2(reg_lamda)))
    model.add(layers.MaxPooling2D(pool_size=(polling_size, polling_size)))

    for i in range(nblocks - 1):
        model.add(
            layers.Conv2D(nfilters,
                          filter_size,
                          activation='relu',
                          kernel_regularizer=keras.regularizers.l2(reg_lamda)))
        model.add(layers.MaxPooling2D(pool_size=(polling_size, polling_size)))

    model.add(layers.Flatten())

    for ndense in dense:
        model.add(
            layers.Dense(ndense,
                         activation='relu',
                         kernel_regularizer=keras.regularizers.l2(reg_lamda)))

    model.add(layers.Dense(8, activation='softmax'))

    return model
```

Code 3. Python code used for defining a function that returns a keras convolutional neural network model for predicting the pitch of cholesteric liquid crystals.

```python
import tensorflow as tf
from tensorflow.keras import layers
from tensorflow import keras

def keras_model_for_predicting_e7_temperature(nblocks=3, nfilters=8,
                                              filter_size=4, polling_size=3,
                                              dense=[32, 16], reg_lamda=0.005):
    """
    Returns a CNN used for predicting the temperature of E7 liquid crystals.

    Parameters:
    nblocks: number of convolution and max-pooling blocks.
    nfilters: number of convolution filters in each convolution block.
    filter_size: size of the convolution filters.
    polling_size: size of the pooling filters.
    dense: a list with the number of neurons in each dense layer.
    reg_lamda: L2 norm for regularization.

    Returns:
    A tf.keras model.

    """
    model = tf.keras.Sequential()
    model.add(
        layers.Conv2D(nfilters,
                      filter_size,
                      activation='relu',
                      input_shape=(510,511,1,),
                      kernel_regularizer=keras.regularizers.l2(reg_lamda)))
    model.add(
        layers.Conv2D(nfilters,
                      filter_size - d_filter_size,
                      activation='relu',
                      kernel_regularizer=keras.regularizers.l2(reg_lamda)))
    model.add(layers.MaxPooling2D(pool_size=(polling_size, polling_size)))

    for i in range(nblocks - 1):
        model.add(
            layers.Conv2D(nfilters,
                          filter_size,
                          activation='relu',
                          kernel_regularizer=keras.regularizers.l2(reg_lamda)))
        model.add(
            layers.Conv2D(nfilters,
                          filter_size - d_filter_size,
                          activation='relu',
                          kernel_regularizer=keras.regularizers.l2(reg_lamda)))
        model.add(layers.MaxPooling2D(pool_size=(polling_size, polling_size)))

    model.add(layers.Flatten())

    for ndense in dense:
        model.add(
            layers.Dense(ndense,
                         activation='relu',
                         kernel_regularizer=keras.regularizers.l2(reg_lamda)))

    model.add(layers.Dense(1, activation='linear'))

    return model
```

Code 4. Python code used for defining a function that returns a keras convolutional neural network model for predicting the temperature of E7 liquid crystals.

```python
from sklearn.metrics import r2_score
from keras import backend as K

def coeff_determination(y_true, y_pred):
    SS_res =  K.sum(K.square( y_true-y_pred ))
    SS_tot = K.sum(K.square( y_true - K.mean(y_true)))
    return (1 - SS_res/(SS_tot + K.epsilon()))

model = keras_model_for_predicting_e7_temperature()

adam = keras.optimizers.Adam(lr=0.001)

model.compile(optimizer=adam, loss='mse', metrics=[coeff_determination, 'mse'])

callback = [keras.callbacks.EarlyStopping(monitor='val_loss', patience=10)]

history = model.fit(X_train,
                    y_train,
                    epochs=500,
                    validation_data=(X_val, y_val),
                    callbacks=callback,
                    verbose=1,
                    batch_size=32)

test_predictions = model.predict(X_test)
r2_score(y_test,test_predictions)
```

Code 5. Example of Python code used training a keras convolutional neural network model for predicting the temperature of E7 liquid crystals. Here X_train, X_val, and X_test represent respectively the image files in the training, validation and test sets. While y_train, X_val, y_val represent respectively the temperature values of the training, validation and test sets.



\end{document}